\documentclass[12pt,preprint]{aastex}

\usepackage{epstopdf}

\slugcomment{ApJ accepted February 9, 2011}

\newcommand\WISE{\emph{WISE}}

\begin{document}

\DeclareGraphicsExtensions{.pdf,.gif,.jpg}

\title{Preliminary Results from NEOWISE: An Enhancement to the \emph{Wide-field Infrared Survey Explorer} for Solar System Science}

\author{A. Mainzer\altaffilmark{1}, J. Bauer\altaffilmark{1}$^{,}$\altaffilmark{2}, T. Grav\altaffilmark{3}, J. Masiero\altaffilmark{1}, R. M. Cutri\altaffilmark{2}, J. Dailey\altaffilmark{2}, P. Eisenhardt\altaffilmark{1}, R. S. McMillan\altaffilmark{4}, E. Wright\altaffilmark{5}, R. Walker\altaffilmark{6}, R. Jedicke\altaffilmark{7}, T. Spahr\altaffilmark{8}, D. Tholen\altaffilmark{7}, R. Alles\altaffilmark{2}, R. Beck\altaffilmark{2}, H. Brandenburg\altaffilmark{2}, T. Conrow\altaffilmark{2}, T. Evans\altaffilmark{2}, J. Fowler\altaffilmark{2}, T. Jarrett\altaffilmark{2}, K. Marsh\altaffilmark{2}, F. Masci\altaffilmark{2}, H. McCallon\altaffilmark{2}, S. Wheelock\altaffilmark{2}, M. Wittman\altaffilmark{2}, P. Wyatt\altaffilmark{2}, E. DeBaun\altaffilmark{9}, G. Elliott\altaffilmark{7}, D. Elsbury\altaffilmark{10}, T. Gautier IV\altaffilmark{11}, S. Gomillion\altaffilmark{12}, D. Leisawitz\altaffilmark{13}, C. Maleszewski\altaffilmark{4}, M. Micheli\altaffilmark{7}, A. Wilkins\altaffilmark{14}}

\altaffiltext{1}{Jet Propulsion Laboratory, California Institute of Technology, Pasadena, CA 91109 USA}
\altaffiltext{2}{Infrared Processing and Analysis Center, California Institute of Technology, Pasadena, CA 91125, USA}
\altaffiltext{3}{Johns Hopkins University, Baltimore, MD USA}
\altaffiltext{4}{Lunar and Planetary Laboratory, University of Arizona, 1629 East University Blvd., Kuiper Space Science Bldg. \#92, Tucson, AZ 85721-0092, USA}
\altaffiltext{5}{UCLA Astronomy, PO Box 91547, Los Angeles, CA 90095-1547 USA}
\altaffiltext{6}{Monterey Institute for Research in Astronomy, Monterey, CA USA}
\altaffiltext{7}{Institute for Astronomy, University of Hawaii, 2680 Woodlawn Drive, Honolulu, HI 96822 USA}
\altaffiltext{8}{Minor Planet Center, Harvard-Smithsonian Center for Astrophysics, 60 Garden Street, Cambridge, MA 02138 USA}
\altaffiltext{9}{Dartmouth College, Hanover, NH 03755 USA}
\altaffiltext{10}{Notre Dame High School, 13645 Riverside Drive, Sherman Oaks, CA 91423 USA}
\altaffiltext{11}{Flintridge Preparatory School, 4543 Crown Avenue, La Canada, CA 91101 USA}
\altaffiltext{12}{Embry-Riddle Aeronautical University, 600 S. Clyde Morris Blvd., Daytona Beach, FL 32114 USA}
\altaffiltext{13}{Goddard Space Flight Center, Greenbelt, MD 20771 USA}
\altaffiltext{14}{University of Maryland, College Park, MD 20742 USA}

\email{amainzer@jpl.nasa.gov}

\begin{abstract}

The \emph{Wide-field Infrared Survey Explorer} has surveyed the entire sky at four infrared wavelengths with greatly improved sensitivity and spatial resolution compared to its predecessors, the \emph{Infrared Astronomical Satellite} and the \emph{Cosmic Background Explorer}.  NASA's Planetary Science Division has funded an enhancement to the \WISE\ data processing system called ``NEOWISE" that allows detection and archiving of moving objects found in the \WISE\ data.  NEOWISE has mined the \WISE\ images for a wide array of small bodies in our Solar System, including Near-Earth Objects (NEOs), Main Belt asteroids, comets, Trojans, and Centaurs.  By the end of survey operations in February 2011, NEOWISE identified over 157,000 asteroids, including more than 500 NEOs and $\sim$120 comets.  The NEOWISE dataset will enable a panoply of new scientific investigations.  

\end{abstract}

\section{Introduction}
A full understanding of the history of our Solar System requires study of its most primitive and dynamic members, the asteroids and comets. These different classes of small bodies are related; recent models of planetary migration, small body collisional history, and the role that primitive bodies have played in shaping our planet's evolution predict that these signatures are imprinted on the numbers, distribution, transport and physical properties of these objects \citep{Gomes, Tsiganis, Morby}.  Only through detailed physical studies of all the major populations of our Solar System's small bodies can their origins, evolution and fate be learned. 

To date, the most prolific surveys to discover and characterize asteroids and comets have been conducted at visible wavelengths \citep[e.g., the Catalina Sky Survey, the Lincoln Near-Earth Asteroid Research Program, the Near Earth Asteroid Tracking Program, and Spacewatch - ][]{Larson,Stokes,Helin,McMillan}.  The number of \emph{in situ} missions to primitive bodies is necessarily small \citep[the \emph{Near-Earth Asteroid Rendevous} mission, \emph{Giotto, Galileo, Deep Space 1, Deep Impact, Hyabusa, Dawn, Stardust, Rosetta}][]{Cheng, Oberst, Reinhard, Thomas, AHearn, Fujiwara, Russell, Brownlee, Barucci}.  Radar observations \citep{Ostro} are a powerful means of obtaining detailed physical characterization remotely, but the objects must pass quite close to Earth. Hence, although approximately 7600 near-Earth objects (NEOs), hundreds of thousands of Main Belt asteroids (MBAs) and $\sim$3000 comets have been discovered to date, detailed physical characterizations have been made for only a small fraction of these objects \citep[e.g.,][]{Binzel, Wolters, Harris09, Bhattacharya}.

\subsection{Infrared Properties of Primitive Bodies}
Thermal infrared observations offer a powerful means of characterizing physical properties of primitive bodies, and all-sky infrared surveys offer an opportunity to obtain these data for large numbers of objects in a relatively short time.  Once an orbit has been calculated for an asteroid, its distance from the Earth at any given time is known, allowing the measured flux to be related directly to the flux emitted at the body's surface.  Likewise, the known distance to the Sun allows derivation of the incident solar flux.  The physical surface properties of the asteroid can then be modeled to reproduce the measured emitted flux as a function of the incident radiation \citep{Harris02, Lebofsky_Spencer}.  Diameters derived from infrared measurements have smaller uncertainties than diameters derived from visible-light observations. Thermal flux emitted by an asteroid has a weaker dependency on a body's unknown albedo value ($p_{v}$) than reflected sunlight. With both visible and IR data, the degeneracy between diameter and albedo can be broken, and both quantities can be independently fit.  Current estimates for the range of albedos for near-Earth objects (NEOs) result in a factor of five uncertainty in diameter for a given reflected optical flux \citep{StuartBinzel}, since optical flux is linearly proportional to $p_{v}$. Yet the variation of an asteroid's thermal flux is proportional to $\left(1-q p_{v}\right)^{1/4}$, where $q$ is the phase integral \citep{Lebofsky_Spencer}. Strong constraints on diameter can still be obtained even for objects which have thermal fluxes but no visible measurements.  For example, for a 0.13 km NEO with a well-known orbit observed to $5\sigma$ at 12 $\mu$m at a distance of 0.5 AU from Earth and 1.1 AU from the Sun, the diameter can be determined to within 10\% \citep{Wright, Mainzer11}.   

Because thermal flux is more weakly dependent on albedo than visible flux, infrared surveys are less subject to the bias against low albedo objects than visible light surveys.  Although considerable work has been done to debias visible surveys \citep[e.g.,][]{StuartBinzel, Jedicke, Harris08}, IR surveys offer a complementary approach to determining the size and albedo distributions of small bodies in the Solar System.  

The \emph{Infrared Astronomical Satellite} observed $\sim$1800 asteroids, most in the Main Belt \citep[\emph{IRAS};][]{Tedesco, Matson}.  Ground-based thermal infrared measurements of asteroids have been made of $\sim$100 objects \citep{Delbo}.  The \emph{Spitzer Space Telescope} has undertaken a number of programs to observe Solar System objects, including 700 NEOs \citep{Trilling}.  However, almost all objects observed by \emph{IRAS}, \emph{Spitzer}, and ground-based infrared telescopes have been optically selected, so the bias against low albedo objects is still present, complicating interpretation of the size and albedo distributions derived from these data.  A large-area infrared survey capable of independent detection and discovery of a statistically significant number of objects is needed.  With the \emph{Wide-field Infrared Survey Explorer} \citep{Wright}, such a resource is now available; the \WISE\ bandpasses cover the thermal peak of most Solar System objects out to $\sim$9 AU.  The  \WISE\ data processing pipeline has been augmented to allow identification and discovery of new moving objects independently of previous visible asteroid surveys; these modifications are collectively known as ``NEOWISE".  Since it is capable of independent discovery of new objects, NEOWISE is unique in its ability to carry out a survey that is much less dependent upon albedo than other surveys.  The purpose of this paper is to introduce the community to the rich NEOWISE dataset, including the pipeline that was used to detect moving objects, and to give an overview of the science enabled by NEOWISE.  More detailed results will follow in separate papers.

\subsection{The \emph{Wide-field Infrared Survey Explorer}}
 \WISE\ is a NASA Medium-class Explorer mission designed to survey the entire sky in four infrared wavelengths, 3.4, 4.6, 12 and 22 $\mu$m (denoted $W1$, $W2$, $W3$, and $W4$ respectively) \citep{Wright, Liu, Mainzer}. \WISE\ is producing a multi-epoch image atlas and source catalog that will serve as an important legacy for future research.   The mission's two primary science goals are to find the most luminous galaxies in the entire universe and to find the closest and coolest stars; a secondary goal is to study the nature of Earth's nearest neighbors, the asteroids and comets.  

 \WISE\ consists of a 40 cm telescope in a Sun-synchronous low-Earth orbit.  The telescope scans the sky continuously, pointing near the zenith while a scan mirror freezes the sky on the focal planes for approximately 11 seconds.  While the sky is frozen on the focal plane, an 8.8 sec exposure is taken of the same 47' x 47' field of view in the four \WISE\ bands.  The scan plane precesses by $\sim$1$^\circ$/day.  Each image overlaps by 10\% in the in-scan direction, with 90\% overlap in the cross-scan direction. This orbit allows  \WISE\ to cover nearly every part of the sky a minimum of eight times, with more coverage at the ecliptic poles.   The survey began on 14 January 2010, and full coverage of the sidereal sky was achieved in mid-July, 2010.  The telescope and detectors were maintained at their operating temperatures through the use of a dual-stage solid hydrogen cryostat.  The secondary cryogen tank depleted on 5 August 2010, and the primary tank was exhausted on 30 September 2010.  After that time, the first two channels ($W1$ and $W2$) continued to operate, and a four month survey (known as the NEOWISE Post-Cryogenic Mission) using these two channels only and the same survey strategy was undertaken. The NEOWISE Post-Cryogenic Mission was completed on 1 February 2011.  

Astrometric errors are less than 0.5 arcsec with respect to 2MASS for objects with high signal-to-noise measurements.  The preliminary estimated SNR=5 point source sensitivity on the ecliptic is 0.08, 0.1, 0.85 and 5.5 mJy in the four bands \citep[assuming eight individual exposures per band;][]{Wright}. Sensitivity improves away from the ecliptic due to denser coverage and lower zodiacal background.  For an asteroid with a bolometric albedo of 0.05, the $W3$ sensitivity of 0.85 mJy corresponds to an visible magnitude of 23 \citep{Wright}. However, as described below, the search for new minor planets was carried out using individual exposures.

The  \WISE\ survey cadence yielded excellent repeated coverage of moving objects in the Solar System, with an average of 10 detections of a typical asteroid or comet spaced over $\sim$36 hours.  This is a slightly higher number of detections than the corresponding value of eight for an inertially fixed object on the ecliptic plane because asteroids tend to move in the same direction as the \WISE\ scan plane.    

As of February 2011, \WISE\ detected more than 157,000 moving Solar System objects.  The dataset contains at least 584 NEOs, $\sim$2000 Jupiter Trojans, 18 Centaurs and scattered-disk objects, and $\sim$120 comets.  \WISE\ has also detected a handful of irregular satellites of Jupiter and Saturn, including Iapetus and Phoebe, as well as cometary dust trails and dust bands in the zodiacal cloud.  

\section{NEOWISE: An Enhancement to the Baseline \WISE\ Mission}
Data processing in the baseline \WISE\ mission does not include any provision for identifying previously unknown Solar System objects.  Furthermore, only the coadded Atlas images that combine many exposures covering each point on the sky were planned to be publicly released. Outlier rejection algorithms designed to reduce susceptibility to transient phenomena such as cosmic rays suppress most moving objects from the coadded images.  To facilitate Solar System science with WISE, two enhancements were made to the baseline pipeline; these tasks are collectively called NEOWISE (``Near Earth Object +  \WISE"). 

\subsection{NEOWISE Task 1: Creation of the archive of single-exposure images plus searchable interface}
NEOWISE Task 1 allows researchers to query the \WISE\ data archive for pre-discovery astrometry and physical data for Solar System objects that are discovered after the final data processing is complete.  It also enables users to download the individual single-epoch images of these objects.

Recovery of small bodies in the \WISE\ data archive requires long-term access to the time-tagged, calibrated images and the database of extracted sources from the individual (or single-epoch) \WISE\ exposures.  These are referred to as the Level 1b image and source products.  The Level 1b images are FITS format images in each of the four bands that have instrumental signatures removed (e.g. linearized, droop-corrected, dark-subtracted, flat-fielded, illumination corrected, etc) and that have the reconstructed WCS and photometric zero point information in their headers. The Level 1b extracted source database contains the reconstructed equatorial positions, calibrated brightness in four bands and associated uncertainties, and assorted quality and characterization flagging for sources detected on the Level 1b images.   Latent images and other instrumental artifacts are flagged in the source database.  These products are produced by the Scan/Frame Pipeline component of the \WISE\ Science Data System (WSDS).

The Atlas, Catalog and assorted metadata products are archived for the long-term and served to the research community and public via the online and machine-friendly interfaces of the NASA Infrared Science Archive (IRSA).  An enhanced interface to the IRSA catalog and image query services that allows searches based on Solar System object name (via Horizons name resolution) or orbital elements has been developed as part of NEOWISE Task 1.  

\subsection{NEOWISE Task 2: Rapid Identification and Posting of Previously Unknown Solar System Object Candidates}
NEOWISE Task 2 consisted of an augmentation to the ground data processing system that exploits the \WISE\ observing cadence to identify moving objects by comparing source extractions made from imaging of the same region of sky on successive orbits.  New moving object candidates were reported to the Minor Planet Center (MPC) within ten days of the midpoint of their observations on board the flight system.  The Minor Planet Center serves as the central clearinghouse for NEOWISE astrometry and associates it with all other observations.

\subsection{WMOPS: The \WISE\ Moving Object Processing Software}
Candidate moving object \emph{tracklets} (sets of time-tagged positions) were identified using the \WISE\ Moving Object Processing System (WMOPS).  WMOPS was developed using elements of the PanSTARRS Moving Object Processing System \citep[MOPS;][]{Denneau, Jedicke09} that linked detections of candidate moving objects from multiple exposures covering the same regions of the sky.  \WISE-specific elements of WMOPS included optimizing detection for the \WISE\ survey cadence, adapting to \WISE\ detector and source extractor characteristics, and specialized quality assurance filters and tools to allow vetting of moving objects prior to delivery to the MPC.  

WMOPS was integrated into the general \WISE\ data processing pipeline system and followed \emph{scan/frame} processing.  The \emph{scan/frame} pipeline performed instrumental image calibration (linearization, dark-subtraction, flat-fielding, sky subtraction, etc.) on individual \WISE\ images from one \emph{scan} (where a scan contains $\sim$260 four-band exposures taken during $1/2$ of a \WISE\ orbit), detected and measured position and fluxes for sources in the calibrated images, and flagged sources that may be spurious detections of image artifacts produced by bright stars.  \emph{Scan/frame} processing produced photometrically and astrometrically calibrated single-exposure (L1b) images and extracted source lists.

\begin{figure}
\figurenum{1}
\plotone{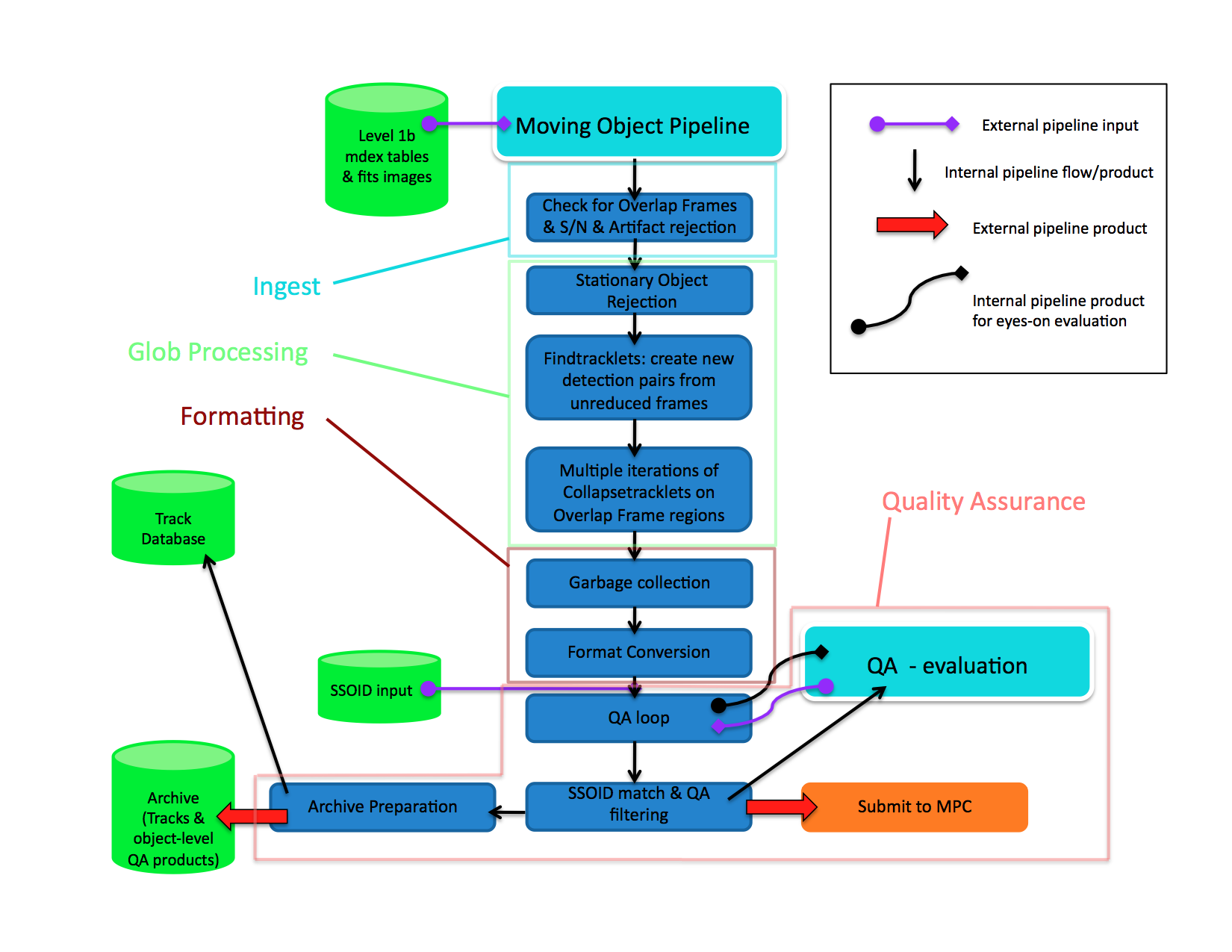}
\caption{\label{fig:flowchart}The flowchart summarizing WMOPS.}
\end{figure}

WMOPS processing steps are illustrated in Figure \ref{fig:flowchart}.  Input was the source lists from $\sim$90-120 scans (spanning approximately 3-4 days).  The input source lists were first filtered to exclude low reliability extractions such as flagged artifacts, very low signal-to-noise ratio sources, and extractions with pathologically large values of the PSF-fitting chi-squared values.  Stationary object rejection (SOR) was then performed by positionally associating the filtered source lists from overlapping images and identifying pairs of detections that do not move by more than 2 arcseconds between adjacent exposures, and eliminating these detections from further processing.  After SOR, the remaining moving object candidate detections are processed by the \emph{findtracklets} routine that creates pairs of detections within a region of sky and correlates each pair's velocity and direction of motion, along with their sky location.  The pair lists are stored in a database and then fed through the \emph{collapsetracklets} routine that creates chains of pair-correlations by matching pairs according to velocity and position within a specified uncertainty threshold at each point.  Candidate tracklets were required to have a minimum of five independent detections to be considered reliable.

\begin{figure}
\figurenum{2}
\plotone{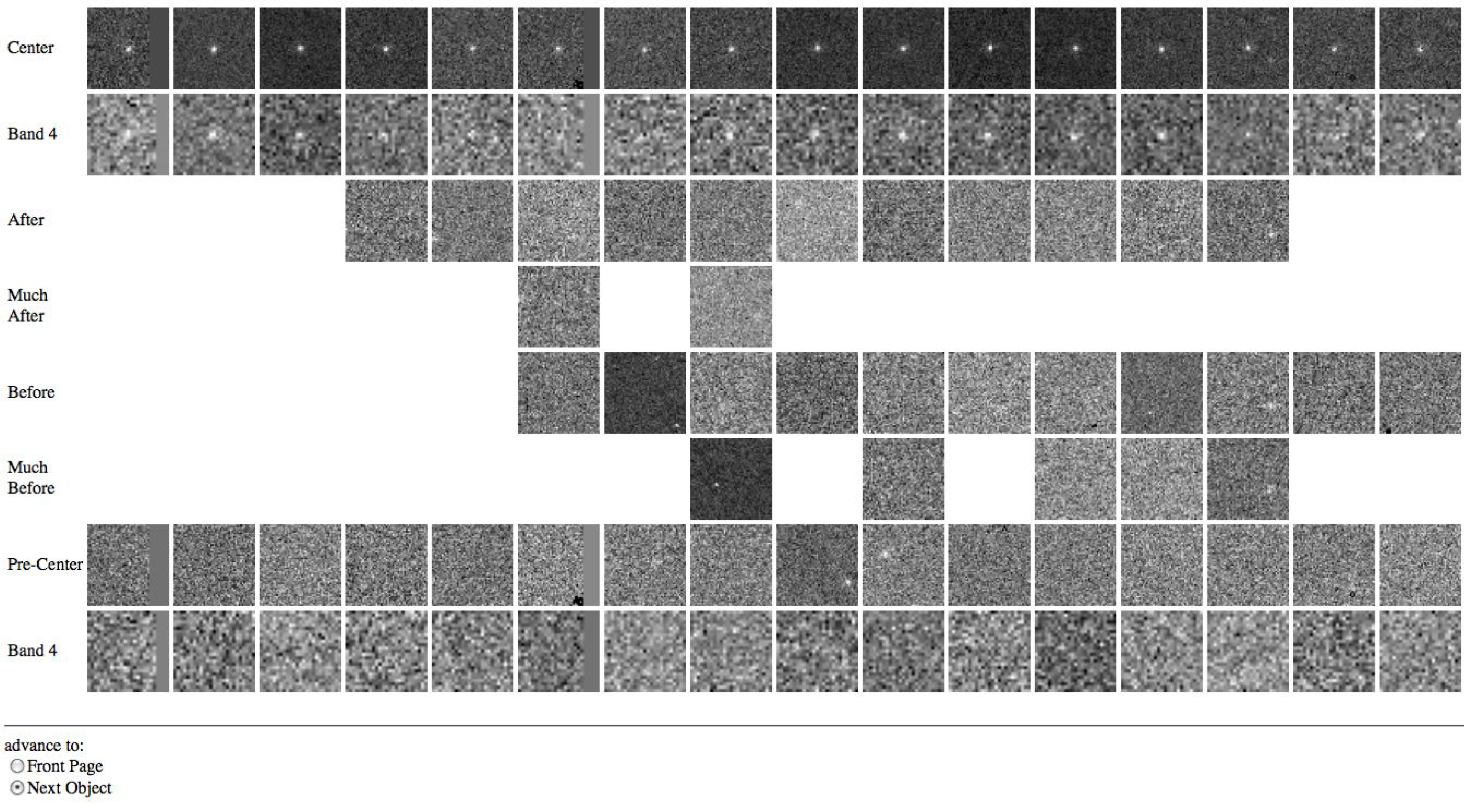}
\caption{\label{fig:tracklet}A sample of the quality assurance pages used to validate a tracklet for a previously unknown moving object.  The top two rows show cutouts of the images in bands $W3$ and $W4$, respectively; the following rows show $W3$ images of the same inertial position on the sky before and after the exposures containing the target.  These rows allow verification that the source observed is not inertially fixed.  The final two rows show the exposures immediately preceding the top row; this allows verification that the transient source observed in the top frame is indeed a moving astronomical object and not an artifact fixed in pixel space but not in time.}
\end{figure}

WMOPS efficiently detected all moving objects whether they were previously known or unknown.  Approximately 80\% of the tracklets were previously known Solar System objects and were automatically accepted.  The remaining 20\% of tracklets were validated by visual inspection by NEOWISE scientists.  Figure  \ref{fig:tracklet} illustrates the web-based quality assurance tool that enabled rapid assessment of a tracklet's reliability.  Tracklets for known objects and reliable new detections were reported to the MPC.  

WMOPS was run approximately every four days to ensure that the accumulated position uncertainties for the fastest moving objects such as NEOs remained small enough to enable efficient recovery by ground-based observers.  Tracklets from each WMOPS run were sent to the MPC for publication and linkage to other observations.  Objects that the MPC determined to have a high probability of being NEOs or comets were posted to the NEO confirmation page (http://www.minorplanetcenter.org/iau/NEO/ToConfirm.html) to be followed up by the world-wide network of professional and amateur astronomers.  Ground-based follow-up was essential for NEOs in particular because the \WISE\ observational arcs, while generally long enough to assign a designation to new objects, do not span a sufficient timebase to allow recovery at subsequent apparitions.  Ground-based follow-up within two to four weeks allows well-determined orbits to be computed.  Objects that are confirmed to be NEOs or comets are published by the MPC via the Minor Planet Electronic Circulars (MPECs) or IAU Circulars.  The astrometry and computed orbits of slower moving objects were published on a monthly basis in the MPC's Minor Planet Orbit Supplement.  

\subsubsection{Range of Motion Limits}
The minimum tracklet length validation requirement and the minimum sky-plane projected velocity imposed by the SOR defines the range of motion to which WMOPS is sensitive.  Parallax motions below 2 arcsec over 90 minutes (approximate time between \WISE\ coverages near the ecliptic) correspond to a distance of roughly 70 AU.  The limits of five detections imposes an upper limit in that objects may outpace the survey cadence before the minimum number of detections can be obtained.  Motions greater than the upper limits perpendicular to the scan direction are roughly 3.25$^\circ$/day, considering both an irregular cadence pattern as well as an evenly progressing survey.  The maximum speed of an object detected by WMOPS was 3.3$^\circ$/day, but the component perpendicular to the scan direction was 3.22$^\circ$/day.

\section{Results}
Figures 4a and 4b show the distribution of orbital elements of all the Solar System objects identified by NEOWISE to date, including NEOs, MBAs, comets, Trojans and Centaurs. By February 2011, more than 252,000 tracklets were reported to the MPC during the cryogenic and post-cryogenic missions, representing observations of $\sim$157,000 unique minor planets.  

\subsection{Near-Earth Objects}
NEOs are thought to be a transient population, persisting in their orbits for only a few million years before being ejected into the outer Solar System, crashing into the Sun, or colliding with other bodies \citep{Gladman}. Therefore, they must be continually resupplied from more permanent source reservoirs such as the Main Belt and comets.  Recent models have improved our understanding of the transport mechanisms and sources of NEOs \citep[e.g.,][]{Bottke} based on observations of their size frequency distribution (SFD).  However, current NEO SFD models rely on the absolute magnitude ($H$) through an assumed albedo to derive size.  By observing and discovering a sample of NEOs at thermal infrared wavelengths, we can obtain an improved computation of the SFD because 1) IR-derived diameters are intrinsically more accurate; 2) we can obtain albedos and further constrain diameters for all individual objects that have measured optical magnitudes; and 3) an IR survey is less biased against low albedo objects.  The improved SFD offered by NEOWISE will allow us to test NEO population models, to study the collisional histories of NEO and MBA populations, as well as to examine how SFD and albedo vary with asteroid collisional families.  

Our understanding of the hazard posed by NEOs also depends on knowledge of their SFD, and efforts to plan mitigation strategies depend on the physical properties of NEOs. To the extent that albedo can be used as a proxy for asteroid composition, it will be possible to derive an approximate density, mass, and estimated impact energy for previously known and new objects observed by WISE.  

As of February 2011, NEOWISE has identified more than 584 NEOs, including discovery of 135 previously unknown NEOs.  Figure \ref{fig:NEOhist} shows the preliminary distribution of $H$ magnitudes for all NEOs known to have been observed by NEOWISE during the cryogenic mission.  The distribution peaks at $H$$\sim$18, corresponding to diameters between $\sim$470 and 1500 m for albedos between 0.05$-$0.5 ($H$ values for previously known objects were taken from the MPC observation file; $H$ values for newly discovered objects were obtained by visible follow-up observations from a worldwide network of amateur and professional observers).  The survey cadence includes many repeated coverages at the ecliptic poles, regions which most ground-based surveys preferentially avoid. NEOWISE is thus sensitive to objects at extreme declinations, which are more likely to have high orbital inclinations or have orbits that bring them very close to the Earth (Figure \ref{fig:iNEO}). Furthermore, unlike ground-based telescopes, the survey is unimpeded by daytime, weather, or variable seeing, so it is easier to debias \citep[c.f.][]{Jedicke}.  However, of the 135 new NEOs discovered, $\sim$15 NEOs were designated by the MPC but received no optical follow-up either due to their optical faintness or unavailability of suitable ground-based facilities.  Due to the  $\sim$36 hour arc length typical of most NEOWISE detections, ground-based follow-up was necessary in most cases to obtain a secure orbit for the newly discovered NEOs.  An additional $\sim$15 NEO candidates were placed on the MPC's NEO Confirmation Page but received neither designations nor follow-up.  A radiometric diameter can be determined for these objects (subject to the uncertainties in Sun-object-observer distances due to orbital uncertainties), but albedo cannot be computed without optical fluxes.  It is hoped that these objects will be recovered either from archival data or future surveys. Without this, these objects will result in an additional uncertainty in the derived NEO albedo distribution. 

\subsection{Main Belt Asteroids}
NEOWISE has observed over 154,000 MBAs as of February 2011, including $\sim$33,000 new discoveries.  This represents an advance of nearly two orders of magnitude over the \emph{IRAS} Minor Planet Survey \citep{Tedesco02}, allowing a correspondingly more detailed view of the size and albedo distributions throughout the Main Belt. Figure \ref{fig:MBA_hist} shows that the preliminary distribution of $H$ magnitudes for all MBAs observed by \WISE\ by the end of the cryogenic mission peaks at $\sim$15, corresponding to diameters between 1.9$-$5.9 km for albedos between 0.05$-$0.5.  $H$ values were taken from the MPC observation files for previously known objects; however, the $H$ values for the newly discovered objects are approximated from the objects' thermal fluxes since no dedicated visible follow-up for these objects was carried out. It is expected that linkages will eventually be made between visible survey data and WISE observations, resulting in increasing numbers of NEOWISE-discovered objects with well-measured $H$ values and refined orbits.  Figure \ref{fig:mba_multiple_hist} shows the preliminary $H$ magnitude distribution as a function of semimajor axis within the Main Belt.  With thermal infrared fluxes available for all of these Main Belt asteroids, it is now possible to derive diameters and (where optical fluxes are available) albedos.  This will facilitate study of the distribution of sizes and albedos within dynamical families \citep{Hirayama} as has been done with multi-band optical photometry \citep{Parker, Ivezic}.     

The NEOWISE dataset contains observations of asteroids that have been previously associated with one of the 54 dynamical families defined by \citet{NesvornyACM}. Figure \ref{fig:family_histogram} shows the number of objects observed by NEOWISE for each family; for some families, NEOWISE has observed thousands of known members.  The combination of WISE infrared fluxes with multi-band optical photometry should allow new family members to be identified on the basis of their albedos and optical colors as well as their dynamical properties.  \citet{Parker} have shown that it is possible to use Sloan Digital Sky Survey (SDSS) colors to find two to three times more family members that make up ``halos" surrounding existing families.  These are family members that have moved far enough into the background population that they can no longer identified by hierarchical clustering methods, yet they have colors identical to the cores of the families (e.g., Vesta and Eos). In particular, this technique may facilitate the association of the smallest, least massive objects with various families.  These objects are of interest because their small sizes make them more susceptible to the non-gravitational forces that cause the most spread in the orbital elements which are used to compute family ages \citep[c.f.][]{NesvornyACM}.  By incorporating size and albedo distributions derived from infrared fluxes, as well as by finding additional family members using their albedos and colors, the NEOWISE dataset should facilitate an improved understanding of the ages and timing of the breakups that created the present-day Main Belt families.   

\subsection{Comets}
As reservoirs of some of the most pristine primordial material in the Solar System, comets serve as Rosetta stones for understanding the transport and history of volatiles.  Comets sample the most distant regions of the Solar System: the Kuiper Belt just outside of Neptune's orbit, and the Oort Cloud, which may extend halfway to the nearest star.  While we cannot directly observe the earliest stages of comet formation in our own Solar System, the size frequency distribution and orbital distribution of present-day comets preserve a record of the outer solar nebula's mass distribution that can be used to model the collisional and dynamical evolution of comet source regions over the last 4.5 Gy \citep[c.f.][]{Tsiganis,Levison}. The cometary size and orbital element distributions allow a test of models of the origins of Halley family comets (HFCs) and long period comets (LPCs) as well as Jupiter family comets (JFCs).  Low-inclination JFCs are thought to originate in the Kuiper belt and are expected to have smaller sizes than the HFCs and LPCs due to collisional processes and mass loss caused by their numerous passages by the Sun \citep{WeissmanLevison}.  The HFCs and LPCs, believed to originate in the Oort cloud, should have retained their original size distribution \citep{SternWeissman} and should be larger \citep{Meech}.  Further, the distribution of perihelia can set constraints on the number and mass of objects in the Oort cloud \citep{Francis}.  

\WISE\ has observed $\sim$120 comets as of February 2011, of which 20 are new discoveries (including three previously known asteroids on which \WISE\ observed activity; these obejcts were subsequently redesignated as comets).  The NEOWISE survey of comets offers a dataset that is less biased against low-albedo objects than previous optical surveys, and as an all-sky survey, it has covered high declination regions as well as the ecliptic plane with a well-determined set of survey pointings.  Detections of previously known comets were identified by WMOPS, and new discoveries were made by inspecting the $\sim$20\% of tracklets that were not associated with known objects to look for evidence of extended emission.  New discoveries suspected of being comets were posted to the MPC's NEO Confirmation Page. All of the 20 new discoveries received visible follow-up, allowing orbits to be refined beyond the initial orbits computed solely from NEOWISE observations.  As with the other major populations detected by \WISE, infrared fluxes can be used to compute nucleus size when cometary activity does not significantly obscure it.  While the \emph{Spitzer Space Telescope} has undertaken a survey of 100 comets \citep{Fernandez}, they are all JFCs that have been selected by optical surveys.  The NEOWISE survey includes both long and short period comets, as well as independent discoveries, allowing constraints to be set on the ratio of long to short period comets. Of the $\sim$120 comets observed by the end of the cryogenic portion of the mission, 35\% are LPCs. Figures \ref{fig:comets} show the distribution of orbital elements of comets detected by NEOWISE during the cryogenic mission.  Many of the comets  have high inclination orbits; however, the survey cadence resulted in increased coverage at the ecliptic poles, leading to a bias in favor of detecting comets at high inclinations (or closer to Earth).  Of the 20 comets discovered by NEOWISE as of February 2011, approximately one-third have perihelia greater than 4 AU, and the remaining two-thirds are LPCs, four of which have nearly parabolic orbits.  

For objects where significant cometary activity is observed, it is possible to compute particle size distributions, allowing the strength of non-gravitational forces to be determined as a function of particle size.  For comets with sufficiently high fluxes at bands $W1$ (3.4 $\mu$m) and $W2$ (4.6 $\mu$m), gas ratios of CO and CO$_2$ can be computed.  

NEOWISE has also observed four objects that have been targeted by spacecraft for \emph{in situ} visits, 19P/Borrelly (\emph{Deep Space 1}), 9P/Tempel 1 (\emph{Deep Impact}), 103P/Hartley \citep[EPOXI;][]{Bauer} and 67P/Churyumov-Gerasimenko, which the \emph{Rosetta} mission will encounter in 2014. Figures \ref{fig:rosetta} show the \WISE\ views of 67P/Churyumov-Gerasimenko at two apparitions.  A preliminary fit of a blackbody function to the first epoch of observations on 19 January 2010 yields an effective temperature of 179 K using an aperture radius of 22 arcsec.  A similar fit to the second epoch observations on 4 July 2010 yields 155 K (Figure 11c).  These fitted temperatures exceed ambient blackbody temperatures by about 20 K.  It is not clear that that this is due to the dust grain temperature, as there is a silicate emission band near the center of \WISE\ band $W3$ at $\sim$10-12 $\mu$m.  

\subsection{Jupiter Trojans}
Trojans are populations of minor planets that are co-orbital with a planet.  The known Jupiter Trojans (there are also a handful of objects that share orbits with Mars and Neptune) are separated into two clouds of objects that librate around the L4 and L5 Lagrangian points in Jupiter's orbit.  These two points are found 60$^\circ$ of heliocentric ecliptic longitude ahead (also called the leading cloud) and behind (the trailing cloud) the planet, and both clouds possess enough dynamical stability to survive over the age of the Solar System \citep{Marzari}.  Reviews of the Trojan populations can be found in \citet{Jewitt} and separately in \citet{Dotto}.  The Trojans are an important population for understanding the origin and evolution of Jupiter, and they also provide a potential window into the solar nebula at the distances of the giant planets, since they are the only major population that might have survived the dynamical clearing of minor planets in this region during the early formation of the Solar System.  But how and when the Trojans were trapped in Jupiter's 1:1 mean-motion resonance remains unknown.  Whether the capture happened during a very early epoch while the planet was forming or much later during the dynamical clearing phase is still under discussion \citep{Morby, MarzariScholl}.

\begin{figure}
\figurenum{3}
\includegraphics[width=6in]{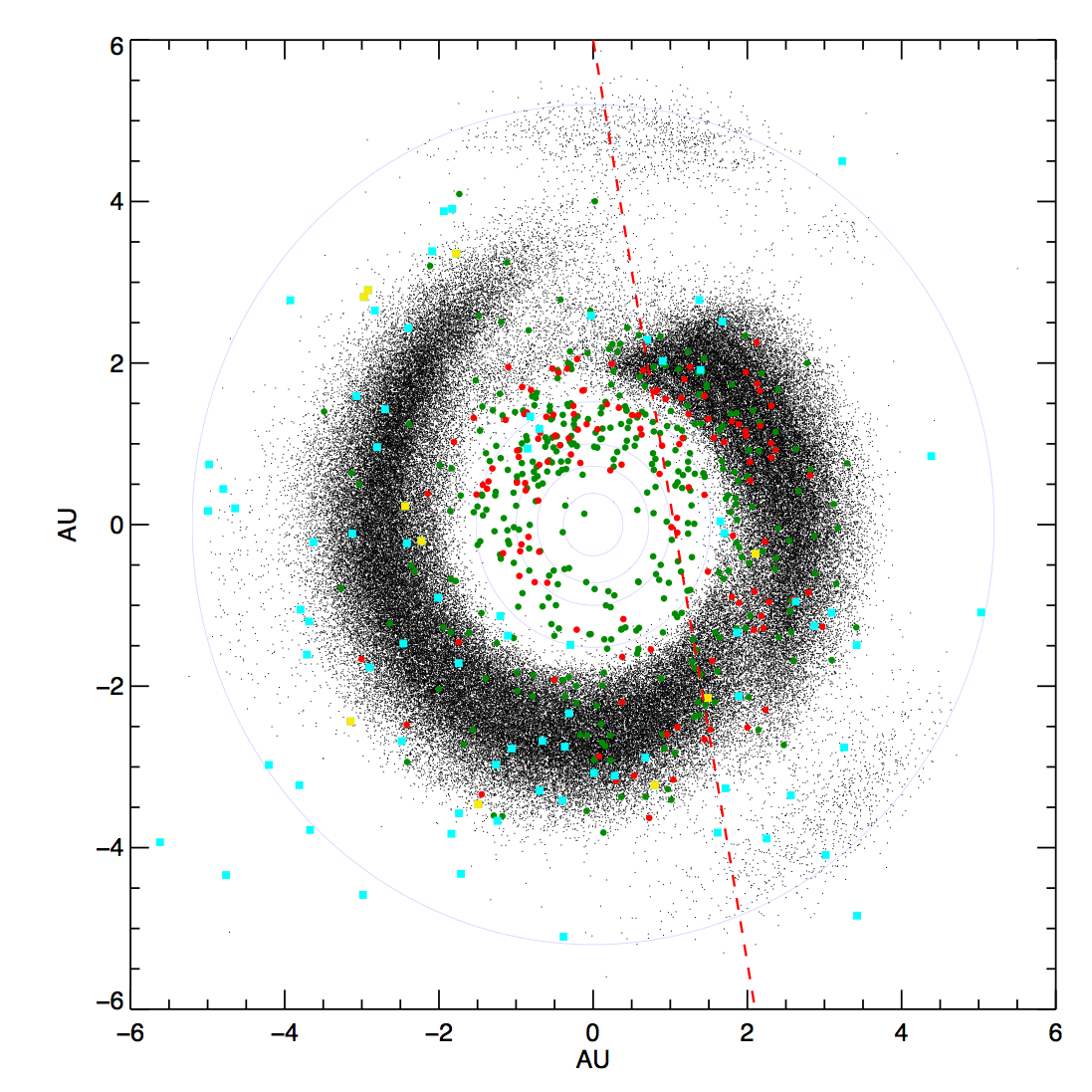}
\caption{\label{fig:xyplot}\WISE\ always surveys near 90$^\circ$ solar elongation; this figure shows a top-down view of the objects detected by NEOWISE as of February 2011 (distances are given in AU). The outermost circle represents Jupiter's orbit; the interior circles represent the terrestrial planets.  Previously known NEOs are shown as green circles; new NEOs discovered by NEOWISE are shown as red circles; previously known comets observed by \WISE\ are shown as cyan squares, and comets discovered by NEOWISE are shown as yellow squares.   All other objects are shown as black points. The drop in density of objects observed near (+2, +2) AU in the figure is due to the exhaustion of the secondary tank's cryogen on 5 August, 2010, resulting in the loss of band $W4$.  The dashed red line indicates the exhaution of the primary tank and the start of the NEOWISE Post-Cryogenic Mission on 1 October 2010; the survey was completed on 1 February 2011.  The drop in detections near (+2, -2) AU in the figure is due to the intersection of the galactic plane with the ecliptic plane; the higher backgrounds and confusion caused by galactic cirrus resulted in the identification of fewer sources.}
\end{figure}

There are currently about 4700 known Jupiter Trojans, and NEOWISE has reported observations of $\sim$1600 of these, increasing the number of thermal emission measurements of this population by more than an order of magnitude compared to earlier surveys \citep{Cruikshank, Tedesco02, Fernandez03, Fernandez09}.  In addition, NEOWISE detected an additional 400 objects that the MPC has designated with Trojan-like orbits, but these have arcs that are too short for their identification to be secure.  \WISE\ observes the sky in scans of constant ecliptic longitude, and as the survey began, the first scans were already in the middle of the leading Trojan cloud.  The scans moved toward the tail of the cloud, resulting in about 20-30\% of the leading cloud closest to Jupiter remaining unobserved during the cryogenic survey (although it was observed about seven months later with bands $W1$ and $W2$ during the post-cryogenic mission).  The survey exited the leading cloud at the end of February, 2010 and almost immediately entered the trailing cloud on the opposite side of the Solar System (Figure \ref{fig:xyplot}).  Observations of the trailing cloud were completed in late May, 2010.  The tail end of the trailing cloud was close to the galactic center when it was observed by NEOWISE, leading to noticeable loss of detection efficiency for less than 10\% of the trailing cloud.  While the two clouds were not observed with perfect uniformity, NEOWISE has provided the most uniform survey of the Jupiter Trojan population to date.  Of the $\sim$1600 objects that have been conclusively identified as Trojans, $\sim$900 were observed the leading cloud and $\sim$700 in the trailing cloud.  This yields a non-debiased ratio of the number of objects in the L4 and L5 cloud of $\sim$1.3.  The 400 probable Trojans are distributed roughly equally among the two clouds.  The H-distribution (Figure \ref{fig:trojan_hist}) shows that the survey's limit is H$\sim$13 (or $\sim$12 km for an assumed albedo of 0.07).

The survey provides a unique opportunity to not only characterize the physical properties of a large fraction of the Trojans, but it also provides a uniform survey of both clouds.  The survey must be debiased in order to test the initial results that show an over-abundance of objects in the leading cloud compared to the trailing cloud.  In addition, a proper vetting of the short arc sample using the on-sky velocity \citep[similar to the method used on the SDSS data by][]{Szabo} and thermal colors is needed to define the Trojan sample.  As with the Main Belt asteroids, NEOWISE detections of new Trojan candidates may be linked to other observations in the future, allowing refinement of their orbits.

\subsection{Unusual Objects and Outer Solar System Objects}
\WISE's wavelengths are not long enough to detect large numbers of objects in the outer Solar System via their thermal emission. Nonetheless, several dozen objects were observed and discovered in this region with the NEOWISE pipeline.  A number of unusual objects with high inclinations have been discovered by NEOWISE such as 2010 OA101, an outer Main Belt Asteroid with an inclination of 84$^\circ$,  and 2010 CR140, an object with a semi-major axis of 5.6 AU and an inclination of 74.7$^\circ$. 2010 CR140 may represent an object in transition between the Jupiter Trojan and Centaur populations.  In total, eleven objects were observed in the outer Solar System with inclinations larger than 80$^\circ$. Figure \ref{fig:outerSS} shows an a-e plot of WISE-observed objects, highlighting the outer Solar System.  \WISE\ has observed $\sim$20 Centaurs to date, including five new discoveries.  

\subsection{Lightcurve Studies and Thermophysical Modeling}
By obtaining an average of ten observations spaced uniformly over $\sim$36 hours for each Solar System object, the \WISE\ cadence lends itself to the study of small body lightcurves.  Ground-based observers are interrupted by daytime, clouds, and seeing variations; however, \WISE\ is able to provide an extremely uniform, stable platform with regular observations spanning days or even, in a few cases, weeks.  An example of this type of long-period lightcurve for a near-Earth asteroid is shown in Figure \ref{fig:lightcurve}.  This makes the \WISE\ dataset unique in comparison with ground-based lightcurve surveys that are generally limited to studying rotational periods faster than $\sim$1 day \citep{Masiero,Pravec08,Pravec06}.  The \WISE\ observations offer a unique window into long-period lightcurve studies, and with suitable debiasing, the fraction of objects with rotational periods longer than a day may be determined.  Figure \ref{fig:num_obs} shows the distribution of the number of observations for objects detected by WMOPS; on average, a typical Solar System object was detected every three hours by NEOWISE.

Binary asteroids have been found via photometric observations, direct imaging (ground-based and space-based), adaptive optics, and radar.  Binary systems provide a useful laboratory because they allow masses to be determined; when combined with infrared diameters, density and porosity may be computed.  It is possible to search for binary systems in the \WISE\ dataset by searching for objects with lightcurve amplitude variations $\gtrsim$1.5 magnitudes; such objects are the most likely to be binary systems \citep{Chandrasekhar,Holsapple}.  An example of a Main Belt asteroid displaying the large-amplitude lightcurve indicative of a candidate binary system is shown in Figure \ref{fig:49137}.  With a period of more than 2.5 days and an amplitude of $\sim$1.5 magnitudes, this object, (49137) 1998 SC35, will require a concerted follow-up observing campaign to confirm its binarity.  Preliminary fits to the individual \WISE\ data points using a thermophysical model assuming a single spherical object and the Near-Earth Asteroid Thermal Model \citep[NEATM;][]{Harris98} yield diameters ranging from 3.8 to 7.5 km; the wide range of diameters is due to rotational effects and is much larger than the measurement uncertainties.  

As with visible data, shape and rotation state can be derived from \WISE\ observations \citep[c.f.][]{Delbo,Wright07}. If sufficient data are available at a range of different phase angles, it is possible to compute the pole orientation and rotation state of an object in addition to its shape model through lightcurve inversion \citep{Kaasalainen,Muller,Durech}.  However, unlike models computed with visible data alone, the infrared fluxes allow the absolute size of the object to be determined as well.  For some \WISE-observed objects, it may be possible to derive an albedo map across an object's surface.  

\subsection{Comet Trails and Dust Bands}
WISE thermal observations are sensitive to extended structures related to the principal mechanisms of dust production in the inner Solar System: comet dust trails and zodiacal dust bands.  Dust trails are narrow (arcminute) structures extending degrees to tens of degrees in the sky, and they consist of low-velocity millimeter to centimeter diameter particle emissions from short-period comets \citep{Sykes,SykesWalker}.  \emph{Spitzer Space Telescope} observations have shown that 80\% of the short-period comets surveyed have trails \citep{Reach}.  A cratering or disruptive collisional event between two asteroids would also result in the formation of a dust trail, though the ejection velocities (compared to trail-forming comet emissions) are expected to be higher, resulting in a broader trail \citep[e.g.][]{Sykes88,Nesvorny}. Over time, gravitational perturbations acting on the trail particles results in the differential precession of their nodes, resulting in the formation of a torus of collisional material extending around the Sun.  Some of these bands have been associated with the breakups of asteroids in the Main Belt that lead to the formation of young asteroid families \citep{Nesvorny03,Nesvorny}. Identifying the dust bands associated with young families will serve to confirm their age and will allow quantitative studies of the dispersal time for the bands \citep{Espy}.  \WISE\ has already detected a number of previously known trails such as that created by Comet 65/P Gunn (Figure \ref{fig:CometGunn}), which was discovered with IRAS and observed by ISO \citep{Sykes86b,Colangeli}.  The \WISE\ image is approximately a hundred times more sensitive than IRAS at 22 $\mu$m and has significantly better spatial resolution.  Future work should allow discovery of more trails that can potentially be associated with either asteroid families or terrestrial meteor streams.  

\section{Conclusions}
With NEOWISE, we have detected over 157,000 minor planets during a year-long survey, including discovery of $\sim$33,000 new minor planets.  We have shown that the WMOPS system is capable of rapid identification of moving objects over a wide range of apparent velocities when provided with appropriately calibrated input images and source detections.  The unique nature of the NEOWISE dataset will facilitate a wide range of small body studies from fields as diverse as population studies, infrared lightcurve shape modeling, thermophysical studies, comets, and dust bands and comet trails.  With close to 100 times more objects observed than its predecessor, \emph{IRAS}, NEOWISE will leave an even greater legacy for future researchers.  

\section{Acknowledgments}

\acknowledgments{This publication makes use of data products from the \emph{Wide-field Infrared Survey Explorer}, which is a joint project of the University of California, Los Angeles, and the Jet Propulsion Laboratory/California Institute of Technology, funded by the National Aeronautics and Space Administration.  This publication also makes use of data products from NEOWISE, which is a project of the Jet Propulsion Laboratory/California Institute of Technology, funded by the Planetary Science Division of the National Aeronautics and Space Administration.  We gratefully acknowledge the extraordinary services specific to NEOWISE contributed by the International Astronomical Union's Minor Planet Center, operated by the Harvard-Smithsonian Center for Astrophysics, and the Central Bureau for Astronomical Telegrams, operated by Harvard University.  We appreciate the contributions of the referee, J. Emery, which greatly improved the manuscript.  We also thank the worldwide community of dedicated amateur and professional astronomers devoted to minor planet follow-up observations. This research has made use of the NASA/IPAC Infrared Science Archive, which is operated by the Jet Propulsion Laboratory, California Institute of Technology, under contract with the National Aeronautics and Space Administration. We thank the Auton Lab of Carnegie Mellon University for use of their libraries, and the PanSTARRS project for MOPS.}


\clearpage

\begin{figure}
\figurenum{4a}
\includegraphics[width=6in]{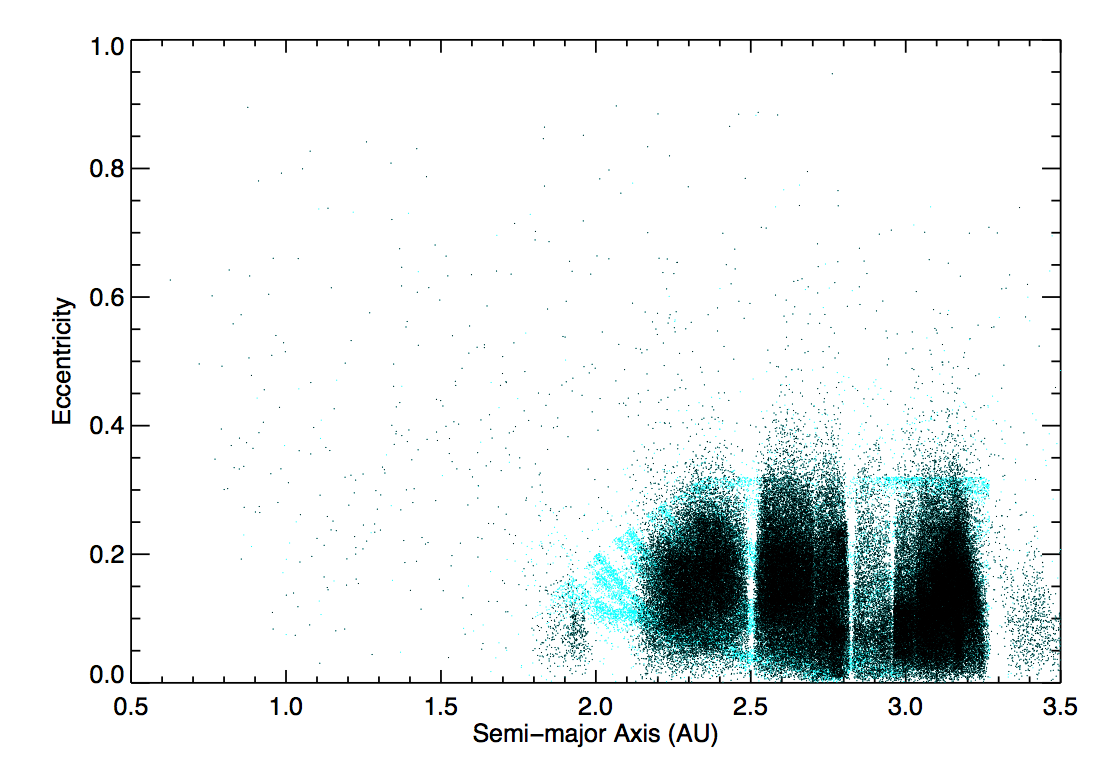}
\caption{NEOWISE has detected objects throughout the Solar System. The orbital elements of all objects that NEOWISE has detected in the inner Solar System with arcs longer than 30 days are shown as black dots; objects with arcs less than 30 days are shown in cyan.  The quantization seen in the cyan points is due to the method by which the Minor Planet Center computes orbits for objects with short observational arcs.}
\end{figure}

\begin{figure}
\figurenum{4b}
\includegraphics[width=6in]{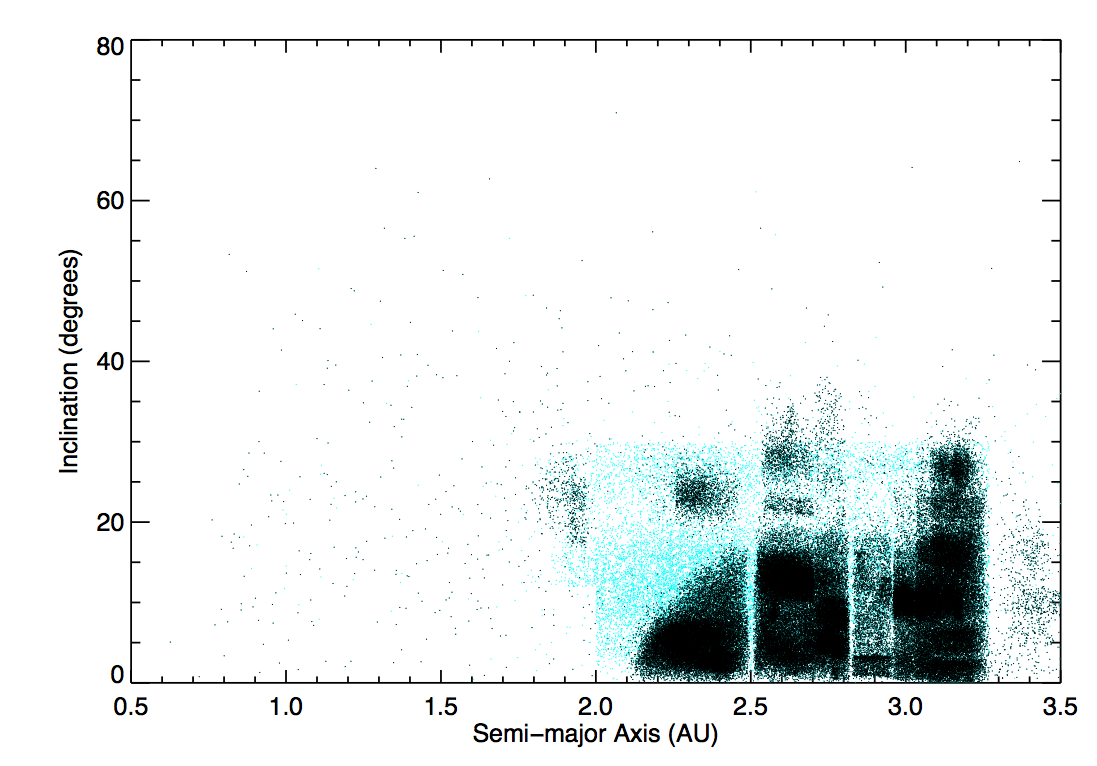}
\figcaption{a-i plot of objects observed by NEOWISE in the inner Solar System; color scheme is the same as in Figure 4a.}
\end{figure}

\begin{figure}
\figurenum{5}
\includegraphics[width=3in]{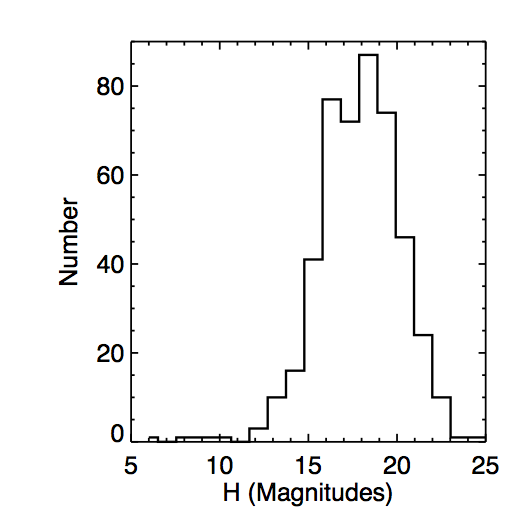}
\caption{\label{fig:NEOhist}The $H$ magnitude distribution for the $>$500 NEOs detected by \WISE\ as of February, 2011.}
\end{figure}

\begin{figure}
\figurenum{6}
\includegraphics[width=3in]{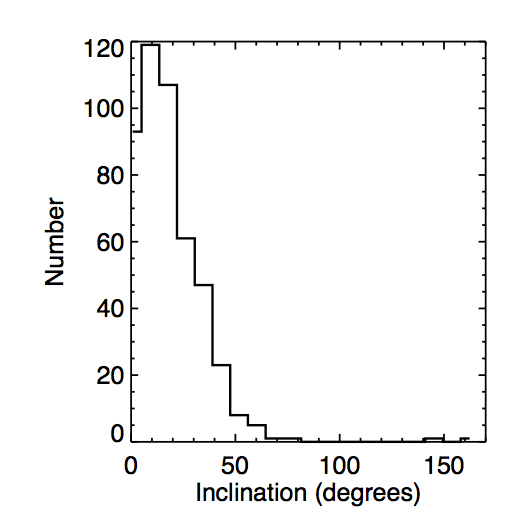}
\caption{\label{fig:iNEO}\WISE's survey cadence results in increased coverage at the ecliptic poles, making it easier to detect NEOs at high inclinations.  The preliminary, non-debiased NEO inclination distribution is shown here for all NEOs observed by WISE.  The dataset must be debiased before conclusions about the NEO inclination distribution can be drawn.}
\end{figure}

\begin{figure}
\figurenum{7}
\includegraphics[width=3in]{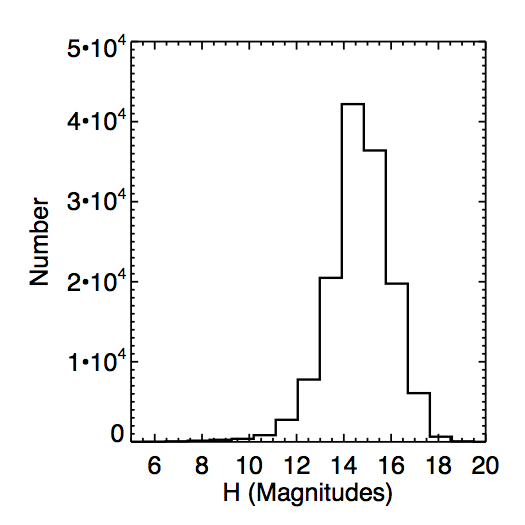}
\caption{\label{fig:MBA_hist}The preliminary distribution of absolute magnitudes of $\sim$150,000 Main Belt asteroids detected by NEOWISE peaks at $H\sim$15; this corresponds to diameters between 1.9$-$5.9 km for albedos of 0.5 to 0.05.}
\end{figure}

\begin{figure}
\figurenum{8}
\includegraphics[width=3in]{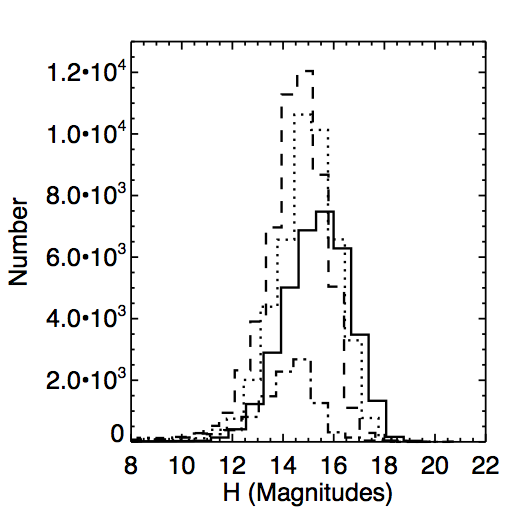}
\caption{\label{fig:mba_multiple_hist}The peak of the Main Belt asteroid $H$ magnitude distributions shifts as a function of semimajor axis.  The plot shows the distributions of four categories of Main Belt asteroids, as defined by Zellner et al. (1979): MBA I (2.06 $<$ a $\leq$ 2.5); MBA II (2.5 $<$ a $\leq$ 2.82); MBA III (2.82 $<$ a $\leq$ 3.27); and MBA IV (3.27 $<$ a $\leq$ 3.65). The solid line represents the histogram of MBA I, the dotted line represents MBA II, the dashed line is MBA III, and the dash-dot line is MBA IV.  The histogram numbers of MBA IV have been multiplied by ten for clarity.}
\end{figure}

\begin{figure}
\figurenum{9}
\includegraphics[width=6in]{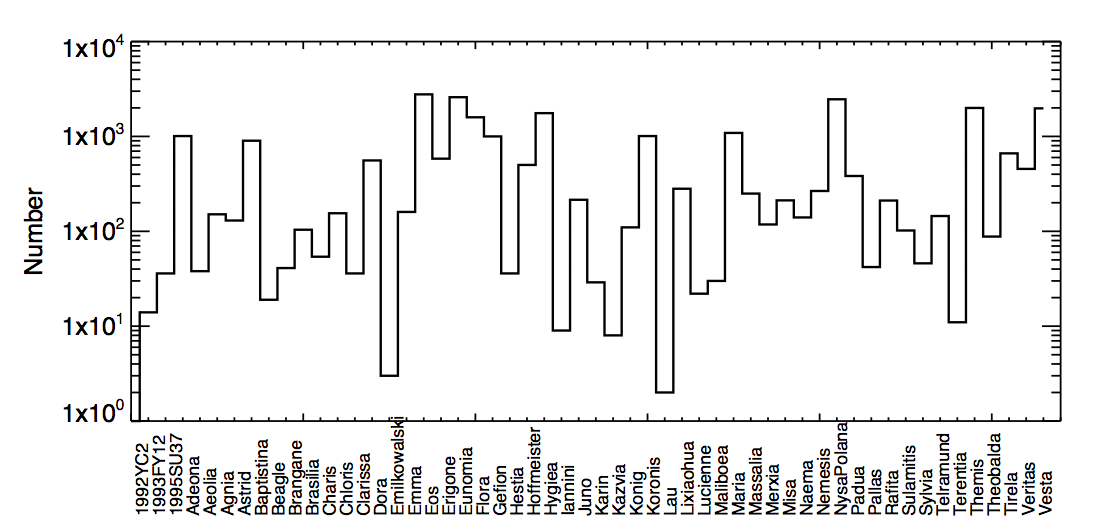}
\caption{\label{fig:family_histogram}Approximately 26,000 of the Main belt asteroids observed by NEOWISE during the cryogenic mission are objects that are known to belong to dynamical families \citet{NesvornyACM}.  In addition to these objects, it is highly likely that other previously unrecognized family members are distributed among the more than 150,000 other MBAs detected by NEOWISE.  These objects may be identified by their IR-derived albedos and optical colors. This will facilitate a wide range of studies aiming to improve the precision of calculations of parent body breakups.}
\end{figure}

\begin{figure}
\figurenum{10ab}
\plottwo {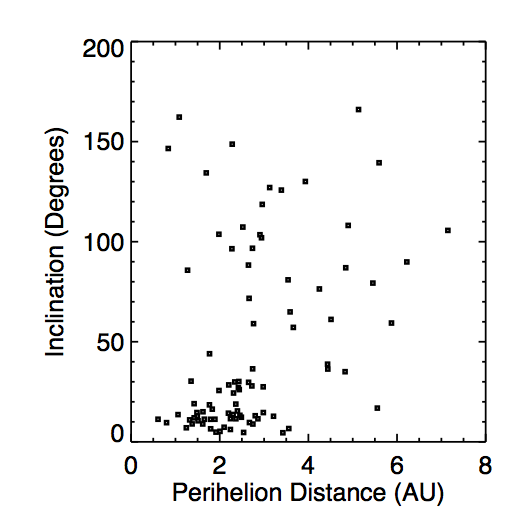}{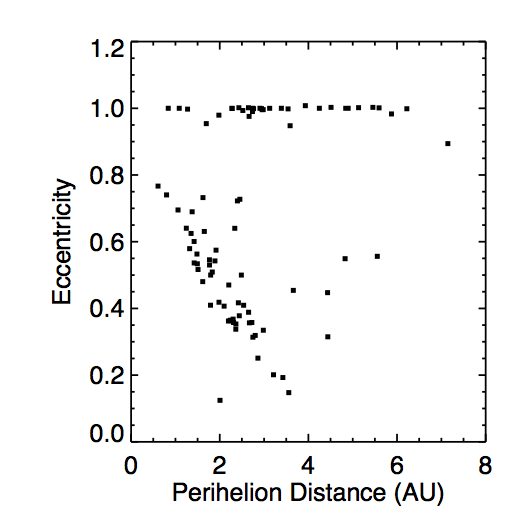}
\figcaption{\label{fig:comets}The preliminary, non-debiased distribution of orbital elements of NEOWISE-detected comets reveals many with high inclinations.}
\end{figure}

\begin{figure}
\figurenum{11abc}
\plottwo{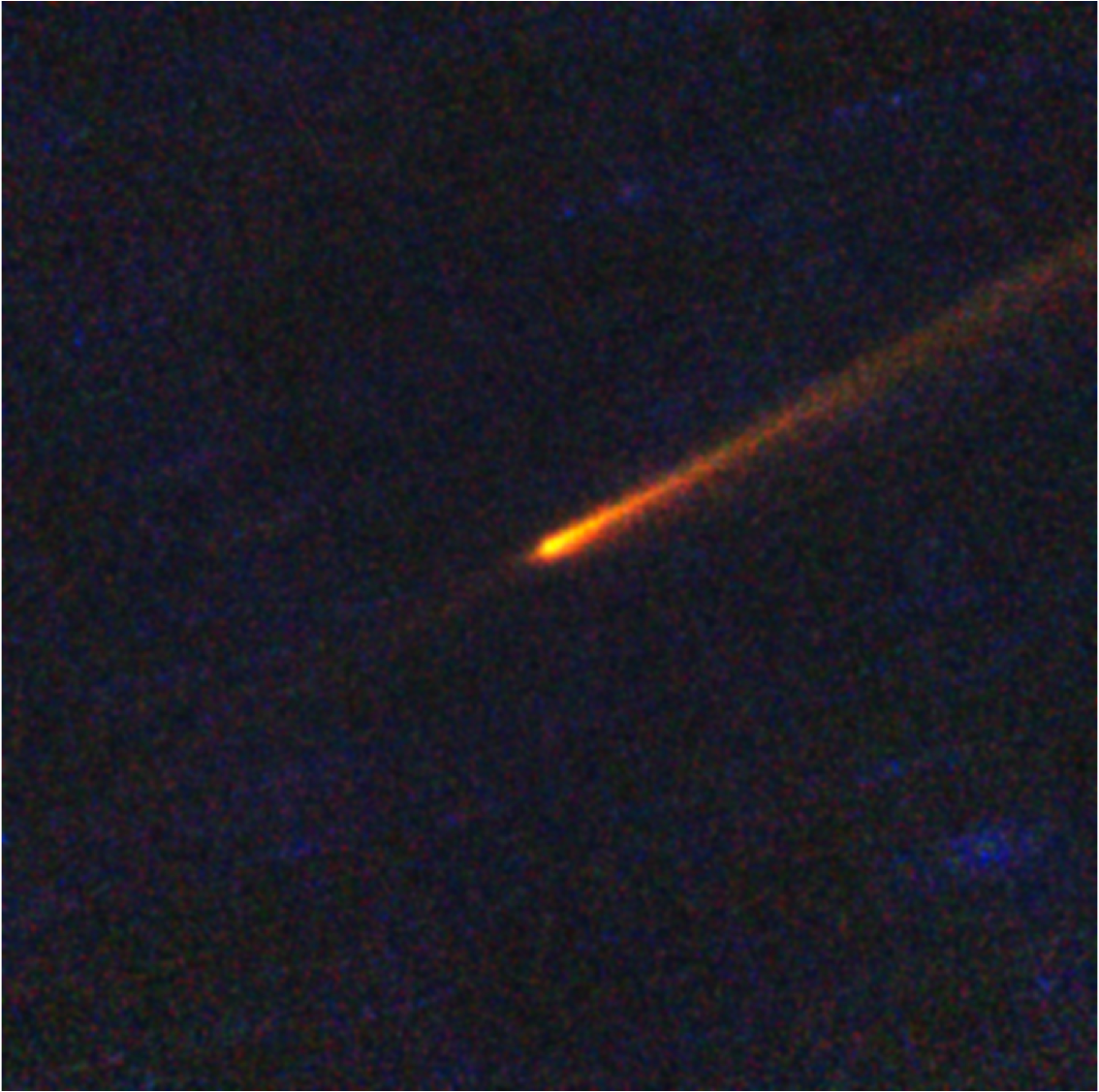}{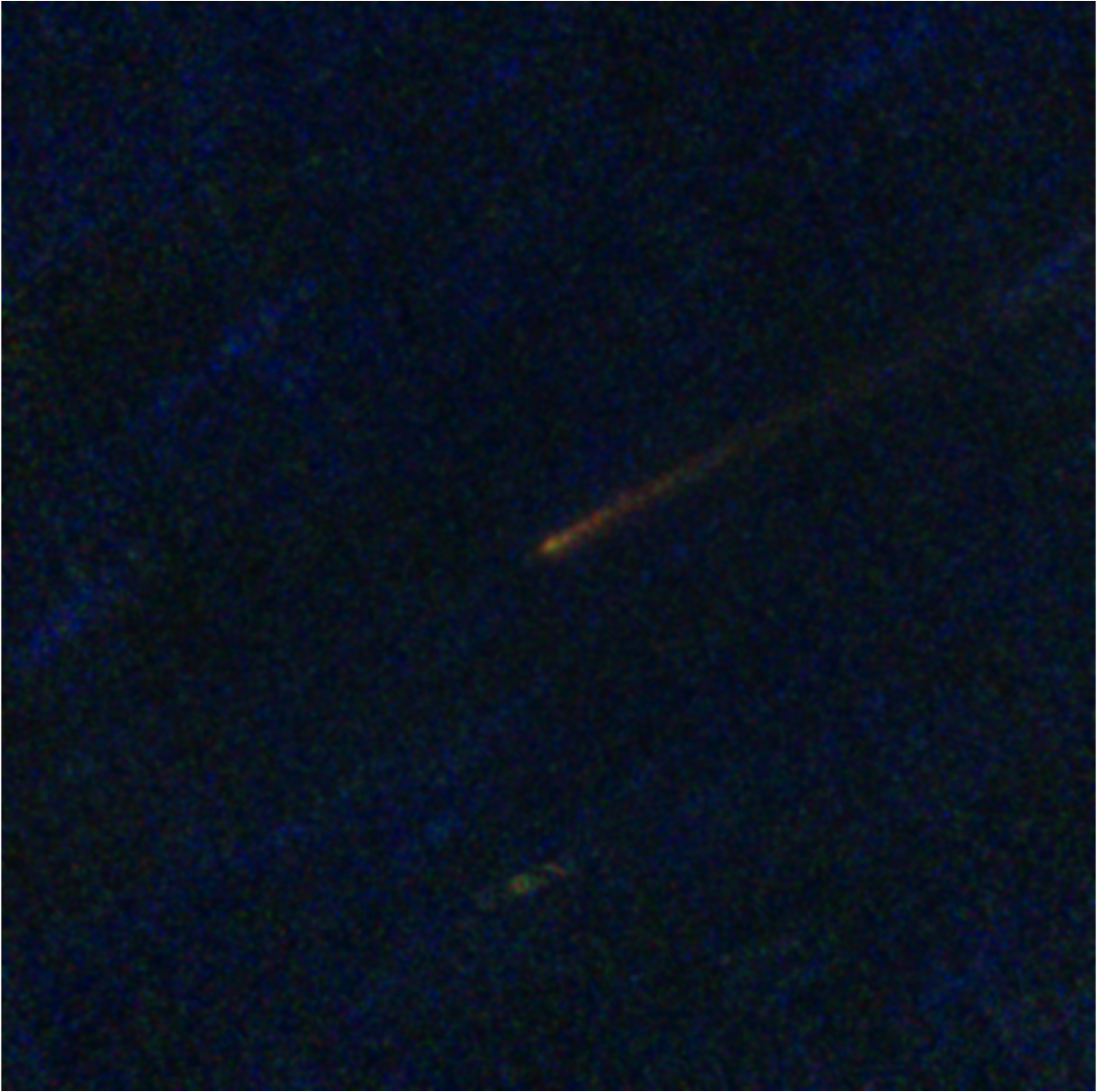}
\includegraphics[width=3in]{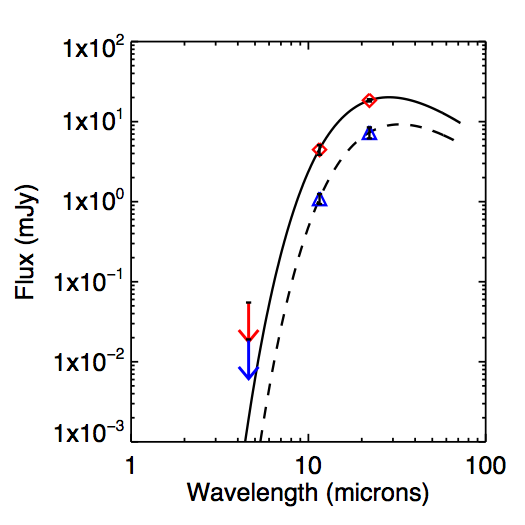}
\caption{\label{fig:rosetta}Rosetta target 67/P Churyumov-Gerasimenko.  11a, b: Three-color \WISE\ images of the comet observed at two epochs: 19 January 2010 (left panel a) at a heliocentric distance $R$ = 3.32 AU and an observer distance $\Delta$=3.14 AU, and 29 June  2010 (right panel b) with $R$ = 4.18 AU, $\Delta$ = 3.97 AU.  In these images, the blue channel is band $W2$ (4.6 $\mu$m), green is band $W3$ (12 $\mu$m), and red is band $W4$ (22 $\mu$m).  11c: Blackbody fits to 67/P for the two epochs of \WISE\ observations.  Preliminary corrections to fluxes were made by interpolating between the blackbody color corrections provided in \citet{Wright}.  The blackbody fit for the first epoch of observations is shown as the solid black line, and the \WISE\ data points are red diamonds; the blackbody fit for the second observational epoch is the dashed black line, and the data points are blue triangles.  67/P was undetected in bands $W1$ and $W2$ in both epochs.}
\end{figure}

\begin{figure}
\figurenum{12}
\includegraphics[width=3in]{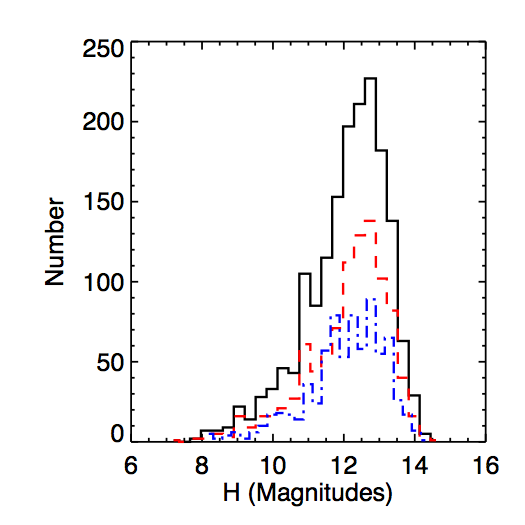}
\caption{\label{fig:trojan_hist}The NEOWISE distribution of observed Trojans peaks at $H\sim$13.  Of the 4237 Trojans known, \WISE\ has observed $\sim$1600 during the one year survey (solid black line), along with $\sim$400 previously unknown Trojans for which orbits are presently poorly constrained.  More Trojans were observed in the L4 cloud that leads Jupiter (red dashed line) than the L5 trailing cloud (blue dash-dot line), despite the fact that the cryogenic portion of the survey ended before the survey of the L4 cloud was complete.}
\end{figure}

\begin{figure}
\figurenum{13}
\includegraphics[width=3in]{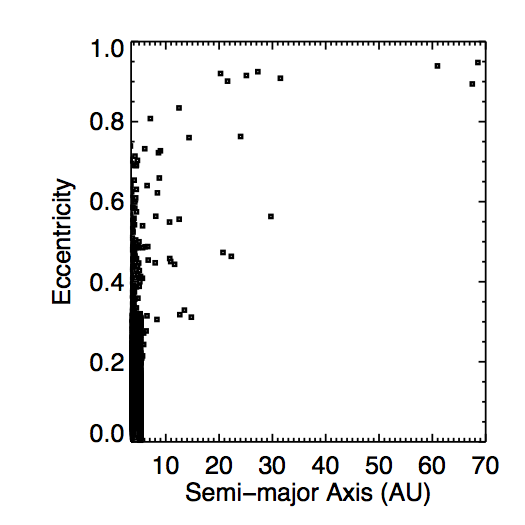}
\figcaption{\label{fig:outerSS}The a-e distribution of objects observed by NEOWISE in the outer Solar System.}
\end{figure}

\begin{figure}
\figurenum{14}
\includegraphics[width=6in]{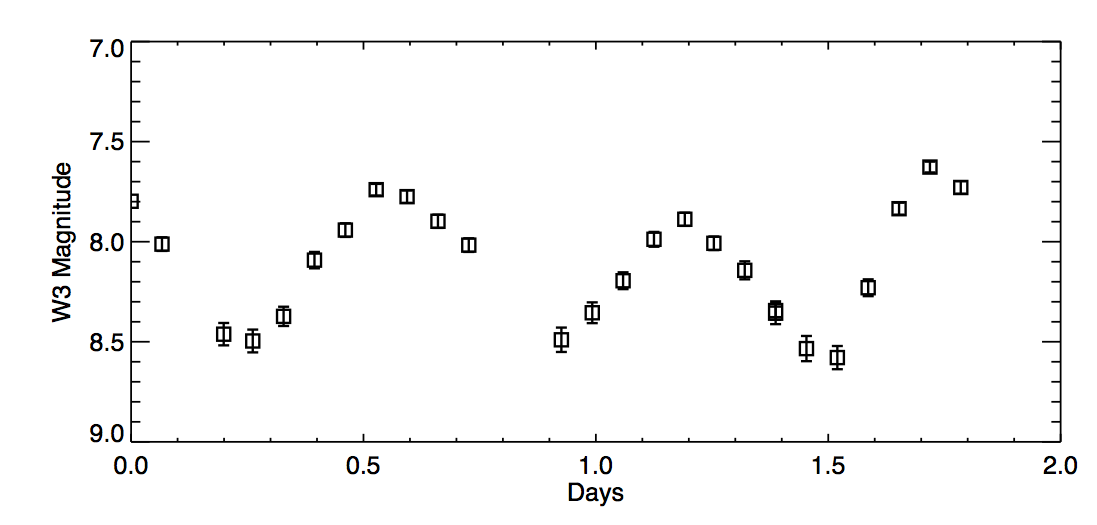}
\caption{\label{fig:lightcurve}The 12 $\mu$m time series of the near-Earth asteroid 230118 (2001 DB3) extends over two days.  The rotational modulation is readily visible due to \WISE's unique observing cadence.} 
\end{figure}

\begin{figure}
\figurenum{15}
\includegraphics[width=3in]{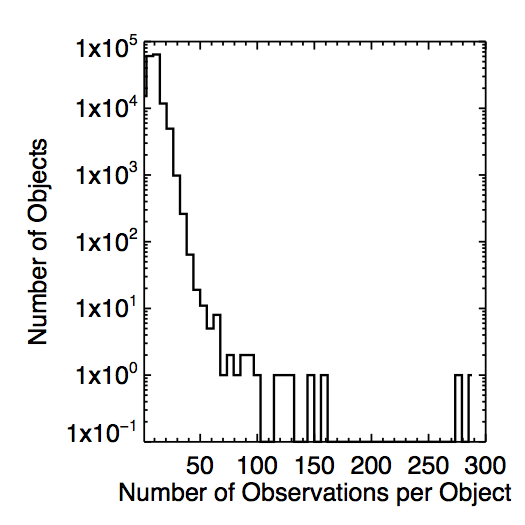}
\caption{\label{fig:num_obs}On average, WMOPS detected most moving Solar System objects $\sim$10-12 times over 1.5 days; however, some NEOs have been observed more than a hundred times.}
\end{figure}

\begin{figure}
\figurenum{16}
\includegraphics[width=3in]{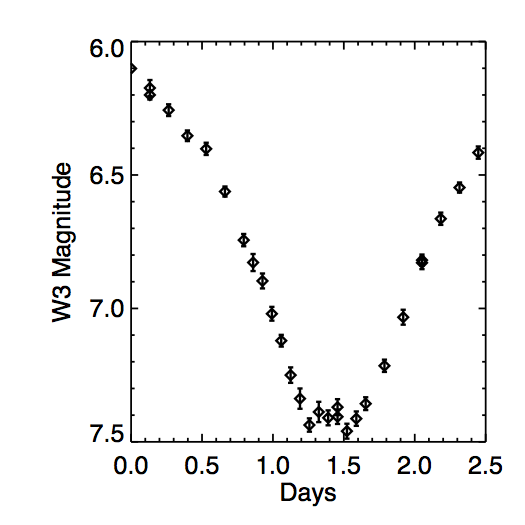}
\caption{\label{fig:49137}The 12 $\mu$m lightcurve of Main Belt Asteroid 49137 extends over 2.5 days and shows an amplitude variation large enough ($\sim$1.5 magnitudes) to indicate that the object may be a binary.  The bottom of the lightcurve shows a flattened region that may indicate the signature of the eclipsing secondary.  The object was detected by WMOPS beginning on Modified Julian Date 55306.640.}
\end{figure}

\begin{figure}
\figurenum{17}
\includegraphics[width=3in]{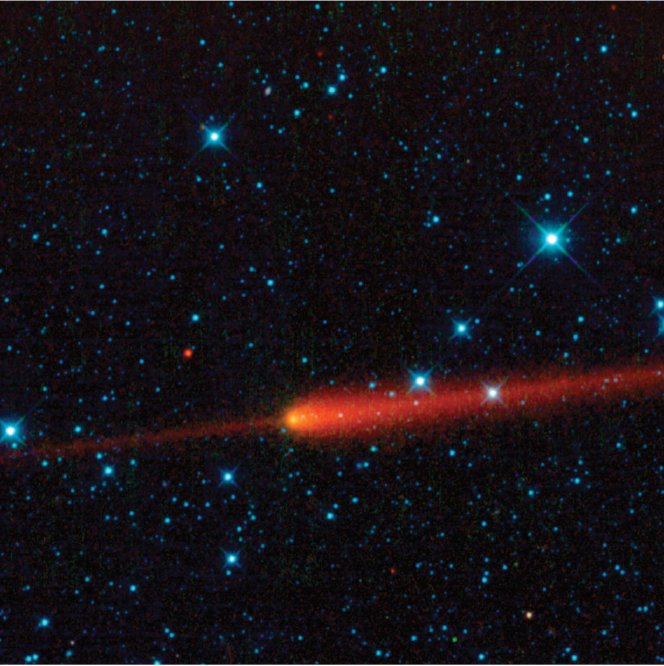}
\caption{\label{fig:CometGunn} The Jupiter-family Comet 65P/Gunn was observed by \WISE\ on 24 April 2010.  This single-frame observation shows 3.4 $\mu$m as blue, 4.6 $\mu$m as green,  12 $\mu$m as orange, and 22 $\mu$m as red.  The nucleus and tail of the comet are clearly visible, along with the trail of debris that has been spread around the comet's orbit.}
\end{figure}

\clearpage

\clearpage
\begin{thebibliography}{}

\bibitem[A'Hearn et al.(2005)]{AHearn}
A'Hearn, M. F.; Belton, M. J. S.; Delamere, W. A.; Kissel, J.; Klaasen, K. P.; McFadden, L. A.; Meech, K. J.; Melosh, H. J.; Schultz, P. H.; Sunshine, J. M.; Thomas, P. C.; Veverka, J.; Yeomans, D. K.; Baca, M. W.; Busko, I.; Crockett, C. J.; Collins, S. M.; Desnoyer, M.; Eberhardy, C. A.; Ernst, C. M.; Farnham, T. L.; Feaga, L.; Groussin, O.; Hampton, D.; Ipatov, S. I.; Li, J.-Y.; Lindler, D.; Lisse, C. M.; Mastrodemos, N.; Owen, W. M.; Richardson, J. E.; Wellnitz, D. D.; White, R. L., 2005, Science, 310, 258

\bibitem[Binzel et al.(2004)]{Binzel}
Binzel, R. P., Rivkin, A. S., Stuart, J. S., Harris, A. W., Bus, S. J., and Burbine, T. H., 2004, Icarus, 170, 254

\bibitem[Bauer et al.(2010)]{Bauer}
Bauer, J., Walker, R.; Mainzer, A.; Masiero, J.; Beck, R.; Masci, F.; Cutri, R.; Wright, E.; A'Hearn, M.; Meech, K.; et al. 2010 CBET 2531

\bibitem[Barucci et al.(2005)]{Barucci}
Barucci, M. A.; Fulchignoni, M.; Fornasier, S.; Dotto, E.; Vernazza, P.; Birlan, M.; Binzel, R. P.; Carvano, J.; Merlin, F.; Barbieri, C.; Belskaya, I., 2005, \aap, 430, 313

\bibitem[Bhattacharya et al.(2010)]{Bhattacharya}
 Bhattacharya, B., Noriega-Crespo, A., Penprase, B., Meadows, V., Salvato, M., Aussel, H., Frayer, D., Ilbert, O., et al. 2010 \apj, 720, 113

\bibitem[Bottke et al.(2002)]{Bottke}
Bottke, W., Morbidelli, A., Jedicke, R., Petit, J.-M., Levison, H., Michel, P., Metcalfe, T., 2002, \icarus, 156, 399

\bibitem[Brownlee et al.(2003)]{Brownlee}
Brownlee, D. E.; Tsou, P.; Anderson, J. D.; Hanner, M. S.; Newburn, R. L.; Sekanina, Z.; Clark, B. C.; Hšrz, F.; Zolensky, M. E.; Kissel, J.; McDonnell, J. A. M.; Sandford, S. A.; Tuzzolino, A. J., 2003, \jgr, 108, 8111

\bibitem[Chandrasekhar(1969)]{Chandrasekhar}
Chandrasekhar, S., Ellipsoidal figures of equilibrium, The Silliman Foundation Lectures, New Haven: Yale University Press, 1969

\bibitem[Cheng(2002)]{Cheng}
Cheng, A., 2002, Asteroids III, W. F. Bottke Jr., A. Cellino, P. Paolicchi, and R. P. Binzel (eds), University of Arizona Press, Tucson, p.351-366

\bibitem[Colangeli et al.(1998)]{Colangeli}
Colangeli, L., Bussoletti, E., Pestellini, C., Fulle, M., Mennella, V., Palumbo, P., Rotundi, A., 1998, \icarus, 134, 35

\bibitem[Cruikshank et al.(1977)]{Cruikshank}
Cruikshank, D., Morrison, D., Pilcher, C., 1977 \apj, 217, 1006

\bibitem[Delb\'{o} et al.(2003)]{Delbo}
Delb\'{o}, M., Harris, A., Binzel, R., Pravec, P., Davies, J. 2003, \icarus, 166, 116

\bibitem[Denneau et al.(2007)]{Denneau}
Denneau, L., Jr., Kubica, J., \& Jedicke, R. 2007, Astronomical Data Analysis Software and Systems XVI, 376, 257  

\bibitem[Dotto et al.(2008)]{Dotto}
Dotto, E., Emery, J., Barucci, M., Morbidelli, A., Cruikshank, D. The Solar System Beyond Neptune, University of Arizona Press, Tucson, 2008, p. 383

\bibitem[\v{Durech} et al.(2010)]{Durech}
\v{Durech}, J., Sidorin, V., \& Kaasalainen, M., 2010 \aap, 513, 46

\bibitem[Espy(2009)]{Espy}
Espy, A.J., Dermott, S.F., Kehoe, T.J., Jayaraman, S., 2009, Planetary and Space Science, 57, 235

\bibitem[Fernandez et al.(2003)]{Fernandez03}
Fernandez, Y., Sheppard, S., Jewitt, D., 2003 AJ 126, 1563

\bibitem[Fernandez et al.(2008)]{Fernandez}
Fernandez, Y., Kelley, M., Lamy, P., Reach, W., Toth, I., Groussin, O., Lisse, C., A'Hearn, M., Bauer, J., Campins, H., Fitzsimmons, A., Licandro, J., Lowry, S., Meech, K., Pittichova, J., Snodgrass, C., Weaver, H. 2008 AGU Spring Meeting P41A-08

\bibitem[Fernandez et al.(2009)]{Fernandez09}
Fernandez, Y., Jewitt, D., Ziffer, J. 2009 AJ, 138, 240

\bibitem[Francis(2005)]{Francis}
Francis, P., 2005, \apj, 635, 1348

\bibitem[Fujiwara et al.(2006)]{Fujiwara}
Fujiwara, A.; Kawaguchi, J.; Yeomans, D. K.; Abe, M.; Mukai, T.; Okada, T.; Saito, J.; Yano, H.; Yoshikawa, M.; Scheeres, D. J.; Barnouin-Jha, O.; Cheng, A. F.; Demura, H.; Gaskell, R. W.; Hirata, N.; Ikeda, H.; Kominato, T.; Miyamoto, H.; Nakamura, A. M.; Nakamura, R.; Sasaki, S.; Uesugi, K., 2006, Science, 312, 1330

\bibitem[Gladman et al.(2000)]{Gladman}
Gladman, B., Michel, P., Froschle, C., 2000, \icarus, 146, 176

\bibitem[Gomes et al.(2005)]{Gomes}
Gomes, R., Levison, H., Tsiganis, K., \& Morbidelli, A., 2005, \nat, 435, 466

\bibitem[Grav et al.(2009)]{Grav}
  Grav, T.; Jedicke, R.; Denneau, L.; Holman, M. J.; Spahr, T.; Pan-STARRS Moving Object Processing System Team, 2009, AAS, 39, 807 

\bibitem[Harris \& Lagerros(2002)]{Harris02}
Harris, A., \& Lagerros, J., 2002, Asteroids III, Eds. Bottke, W., Cellino, A., Paolicchi, P., and Binzel, R., University of Arizona Press, Tucson, p. 205

\bibitem[Harris(2008)]{Harris08}
Harris, A., 2008, \nat, 453, 1178

\bibitem[Harris et al.(2009)]{Harris09}
Harris, A., Mueller, M., Lisse, C., Cheng, A., 2009, \icarus, 199, 86

\bibitem[Harris(1998)]{Harris98}
Harris, A. 1998, \icarus, 131, 291

\bibitem[Helin et al.(1997)]{Helin}
Helin, E., Pravdo, S., Rabinowitz, D., Lawrence, K. 1997, Near-Earth Objects, the United Nations International Conference: Proceedings of the international conference held April 24-26, 1995 in New York, New York, USA. Edited by John L. Remo, 1997. Annals of the New York Academy of Sciences, vol. 822, p. 6

\bibitem[Hirayama(1918)]{Hirayama}
Hirayama, K., 1918, \aj, 31, 185

\bibitem[Holsapple(2001)]{Holsapple}
Holsapple, K., 2001, \icarus, 154, 432

\bibitem[Ivezic et al.(2002)]{Ivezic}
Ivezi$\acute{c}$, $\breve{Z}$, Lupton, R.; Juri$\acute{c}$, M.; Tabachnik, S.; Quinn, T.; Gunn, J.; Knapp, G.; Rockosi, C.; Brinkmann, J., 2002, \aj, 124, 2943

\bibitem[Jedicke \& Metcalfe(1998)]{Jedicke}
Jedicke, R., \& Metcalfe, T., 1998, \icarus, 131, 245

\bibitem[Jedicke et al.(2009)]{Jedicke09}
Jedicke, R., Denneau, L., Granvik, M., \& Wainscoat, R. 2009, Proceedings of the Advanced Maui Optical and Space Surveillance Technologies Conference, held in Wailea, Maui, Hawaii, September 1-4, 2009, Ed.: S. Ryan, The Maui Economic Development Board., p.E43

\bibitem[Jewitt et al.(2004)]{Jewitt}
Jewitt, D., Sheppard, S., Porco, C., Jupiter. The planet, satellites and magnetosphere.  Edited by F. Bagenal, T. Dowling, W. McKinnon.  Cambridge planetary science, Vol. 1, Cambridge, UK: Cambridge University Press, 2004, p. 263

\bibitem[Kaasalainen et al.(2002)]{Kaasalainen}
Kaasalainen, M., Mottola, S., Fulchignoni, M., 2002 Asteroids III

\bibitem[Larsen(2007)]{Larson}
Larsen, S., Near Earth Objects, our Celestial Neighbors: Opportunity and Risk, Proceedings if IAU Symposium 236. Edited by G.B. Valsecchi and D. Vokrouhlicky, and A. Milani. Cambridge: Cambridge University Press, 2007., pp.323-328

\bibitem[Lebofsky \& Spencer(1989)]{Lebofsky_Spencer}
Lebofsky, L., \& Spencer, J., Asteroids II, University of Arizona Press, 1989, p. 128-147

\bibitem[Levison et al.(2009)]{Levison}
Levison, H., Bottke, W., Gounelle, M., Morbidelli, A., Nesvorny, D., Tsiganis, K., 2009, Nature, 460, 364

\bibitem[Liu et al.(2008)]{Liu}
  Liu, F.; Cutri, R.; Greanias, G.; Duval, V.; Eisenhardt, P.; Elwell, J.; Heinrichsen, I.; Howard, J.; Irace, W.; Mainzer, A.; Razzaghi, A.; Royer, D.; Wright, E. L., 2008, SPIE, 7017, 16
  
\bibitem[Mainzer et al.(2011)]{Mainzer11}
Mainzer, A., T. Grav, J. Masiero, J. Bauer, E. Wright, R. M. Cutri, R. S. McMillan, M. Cohen, M. Ressler, P. Eisenhardt, D. Leisawitz 2011 ApJ submitted
  
\bibitem[Mainzer et al.(2005)]{Mainzer}
  Mainzer, A.; Eisenhardt, P.; Wright, E. L.; Liu, F.; Irace, W.; Heinrichsen, I.; Cutri, R.; Duval, V.  2005, SPIE, 5899, 262
  
\bibitem[Marzari \& Scholl(1998)]{Marzari}
Marzari, F., Scholl, H. 1998 \icarus, 131, 41   

\bibitem[Marzari \& Scholl(2007)]{MarzariScholl}
Marzari, F., Scholl, H., 2007 MNRAS, 380, 479
  
\bibitem[Masiero et al.(2009)]{Masiero}
Masiero, J., Jedicke, R., Durech, J., Gwyn, S., Denneau, L., Larsen, J. 2009 \icarus, 204, 145

\bibitem[Matson, D.(1986)]{Matson}
Matson, D., ed. The IRAS Asteroid and Comet Survey, 1986, JPL D-3698 (Pasadena: JPL).

\bibitem[McMillan(2007)]{McMillan}
McMillan, R. S., Near Earth Objects, our Celestial Neighbors: Opportunity and Risk, Proceedings if IAU Symposium 236. Edited by G.B. Valsecchi and D. Vokrouhlicky, and A. Milani. Cambridge: Cambridge University Press, 2007., pp.329-340

\bibitem[Meech et al.(2004)]{Meech}
Meech, K., Hainaut, O., Marsden, B., 2004, Icarus, 170, 463

\bibitem[Morbidelli et al.(2005)]{Morby}
Morbidelli, A., Levison, H., Tsiganis, K., Gomes, R., 2005, \nat, 435, 462

\bibitem[M\"{u}ller et al.(2005)]{Muller}
M\"{u}ller, T., Sekiguchi, T., Kaasalainen, M., Abe, M., Hasegawa, S., 2005, \aap, 347, 355

\bibitem[Nesvorny et al.(2003)]{Nesvorny03}
Nesvorny, D., Bottke, W. F., Levison, H. F., Dones, L. 2003, \apj, 591, 486

\bibitem[Nesvorny et al.(2006a)]{NesvornyACM}
Nesvorny, D., Bottke, W., Vokrouhlicky, D., Morbidelli, A., Jedicke, R., 2006, Proceedings of the International Astronomical Union, Cambridge University Press, p. 289

\bibitem[Nesvorny et al.(2006b)]{Nesvorny}
Nesvorny, D., M. Sykes, D.J. Lien, J. Stansberry, W.T. Reach, D. Vokrouhlicky, 
W.F. Bottke, D.D. Durda, S. Jayaraman, R.G. Walker, 2006, \aj, 132, 582

\bibitem[Oberst et al.(2001)]{Oberst}
Oberst, J., Mottola, S., Di Martino, M., Hicks, M., Burrati, B., Soderblom, L, Thomas, N. 2001, Icarus, 153, 16

\bibitem[Ostro et al.(2007)]{Ostro}
Ostro, S. J., Giorgini, J., Benner, L., 2007, Near Earth Objects, our Celestial Neighbors: Opportunity and Risk, Proc. of IAUS 236. Eds. Valsecchi, G., Vokrouhlicky, D., Cambridge: Cambridge University Press, 143

\bibitem[Parker et al.(2008)]{Parker}
Parker, A.; Ivezi$\acute{c}$, $\breve{Z}$; Juri$\acute{c}$, M.; Lupton, R.; Sekora, M. D.; Kowalski, A., 2008, \icarus, 198, 138

\bibitem[Pravec et al.(2006)]{Pravec06}
Pravec, P., et al. 2006, \icarus, 181, 63

\bibitem[Pravec et al.(2008)]{Pravec08}
Pravec, P., et al. 2008, \icarus, 197, 497

\bibitem[Reach et al.(2007)]{Reach}
Reach, W.T., Kelley, M.S., \& Sykes, M.V., 2007, \icarus 19, 298

\bibitem[Reinhard(1986)]{Reinhard}
Reinhard, R., 1986, \nat, 321, 313

\bibitem[Russel et al.(2007)]{Russell}
Russell, C. T.; Capaccioni, F.; Coradini, A.; de Sanctis, M. C.; Feldman, W. C.; Jaumann, R.; Keller, H. U.; McCord, T. B.; McFadden, L. A.; Mottola, S.; Pieters, C. M.; Prettyman, T. H.; Raymond, C. A.; Sykes, M. V.; Smith, D. E.; Zuber, M. T., 2007, EM\&P, 101, 65

\bibitem[Stern \& Weissman(2001)]{SternWeissman}
Stern, S., \& Weissman, P., 2001, \nat, 409, 589

\bibitem[Stokes et al.(2000)]{Stokes}
Stokes, G., Evans, J., Viggh, H, Shelly, F., Pearce, E., 2000, \icarus, 148, 21

\bibitem[Stuart \& Binzel(2004)]{StuartBinzel}
Stuart, J., \& Binzel, R., 2004, \icarus, 170, 295

\bibitem[Sykes et al.(1986a)]{Sykes}
Sykes, M., et al., 1986, Science, 232, 1115

\bibitem[Sykes et al.(1986b)]{Sykes86b}
Sykes, M., Hunten, D., Low, F., 1986, Adv. Space Res. 6, 7, 67

\bibitem[Sykes(1988)]{Sykes88}
Sykes, M., 1988, \apj, 334, L55

\bibitem[Sykes \& Walker(1992)]{SykesWalker}
Sykes, M., \& Walker, R., 1992, \icarus, 95, 180

\bibitem[Szabo et al.(2007)]{Szabo}
Szabo, G., Ivezic, Z., Juric, M., Lupton, R. 2007 MNRAS, 377, 1393

\bibitem[Tedesco et al.(1988)]{Tedesco}
Tedesco, E., Matson, D., Veeder, G., Lebofsky, L., 1988, Comets to Cosmology: Proceedings of the Third IRAS Conference, Springer Berlin, Heidelberg, Vol. 297, p. 19-26

\bibitem[Tedesco et al.(2002)]{Tedesco02}
Tedesco, E., Noah, P., Noah, M., Price, S. 2002 \aj, 123, 1056

\bibitem[Thomas et al.(1994)]{Thomas}
Thomas, P., Veverka, J., Simonelli, D., Helfenstein, P., Carcich, B., Belton, M., Davies, M., Chapman, C. 1994, 

\bibitem[Trilling et al.(2010)]{Trilling}
Trilling, D., et al.  2010, \aj, 140, 770

\bibitem[Tsiganis et al.(2005)]{Tsiganis}
Tsiganis K, Gomes R, Morbidelli A, Levison, H., 2005, \nat, 435, 459

\bibitem[Weissman \& Levison(1997)]{WeissmanLevison}
Weissman, P., \& Levison, H. 1997 

\bibitem[Wolters et al.(2008)]{Wolters}
  Wolters, Stephen D.; Green, Simon F.; McBride, Neil; Davies, John K., 2008, \nat, 193, 535

\bibitem[Wright et al.(2010)]{Wright}
  Wright, E. L., et al. 2010 \aj, 140, 1868
  
\bibitem[Wright et al.(in prep.)]{WrightNEO}
Wright, E. L., et al. 2010 \aj, in preparation 

\bibitem[Wright et al.(2007)]{Wright07}
Wright, E. L., 2007 astro-ph/0703085

\end{thebibliography}
\end{document}